\documentclass{article}
\usepackage{graphicx}

\usepackage{floatrow}

\usepackage{sidecap} 

\usepackage{lipsum}

\usepackage{blindtext}

\usepackage{geometry} 

\usepackage{setspace} 
\usepackage[font=singlespacing]{caption}

\usepackage[superscript,move]{cite} 

\makeatletter
\newenvironment{figurehere}
{\def\@captype{figure}}
{}

\makeatletter
\renewcommand\@citess[1]{\textsuperscript{[#1]}}

\begin{document}

\begin{titlepage}
\vspace*{\stretch{1.0}}
\begin{center}
\Large\textbf{Immobilization of water drops on hydrophobic surfaces by contact line pinning at non-lithographically generated polymer microfiber rings}\\
\vspace*{\stretch{1.0}}

\large\textit{Peilong Hou, Martin Steinhart$^\ast$}
\vspace*{\stretch{1.0}}

\large{Institut f\"ur Chemie neuer Materialien, Universit\"at Osnabr\"uck, Barbarastr. 7, 49076 Osnabr\"uck, Germany\\
	martin.steinhart@uos.de}
\vspace*{\stretch{1.0}}

\end{center}
\end{titlepage}

\newpage
\begin{abstract}
Water drops used as reaction compartments are commonly immobilized on hydrophilic areas bordered by hydrophobic areas. For many applications, such as the trapping of non-adherent cells, it is desirable to exploit the inertness and the anti-fouling behavior of hydrophobic surfaces as well as their repulsive behavior towards adsorbates in lab-on-chip configurations. However, the immobilization of water drops on hydrophobic surfaces has remained challenging. We report a nonlithographic approach to arrest water drops on hydrophobically modified macroporous silicon (mSi) with perfluorinated surface. Contact line pinning at rings of polystyrene-\textit{block}-poly(2-vinylpyridine) (PS-\textit{b}-P2VP) fibers protruding from the mSi macropores immobilizes water drops when the hydrophobically modified mSi is moved or tilted and prevents dewetting within the PS-\textit{b}-P2VP fiber rings. Without PS-\textit{b}-P2VP fiber rings, water drops readily roll off. The PS-\textit{b}-P2VP fiber rings were prepared by dropping PS-\textit{b}-P2VP solution onto hydrophobically modified mSi. Selective swelling of the P2VP in the thus-formed circular PS-\textit{b}-P2VP films with hot ethanol followed by detachment of the latter yielded hydrophobically modified mSi exhibiting annular areas, in which ruptured PS-\textit{b}-P2VP fibers protruded from the mSi macropores. For example, PS-\textit{b}-P2VP fiber rings with diameters of 6.5 mm and widths of $\sim$0.2 mm immobilize water drops with a volume of 50 $\mu$L.
\end{abstract}

\newpage
\section{Introduction}
Discrete water drops immobilized on substrates have been explored for a diverse range of applications including the trapping of bioactive molecules, nonadherent cells, and microorganisms \cite{MD_Ueda2012}. Moreover, immobilized  water drops were employed for enzymatic reactions \cite{MD_Huebner2009}, synthesis and profiling of enzyme inhibitors \cite{MD_Mugherli2009}, combinatorial discovery of fluorescent pharmacophores \cite{MD_Burchak2011}, high-throughput screenings of embryonic stem cells \cite{MD_Tronser2017}, whole-organism screenings using fish embryos \cite{MD_Popova2018}, as well as for the creation of cell type patterns \cite{MD_Efremov2013}, microdrop-derived hydrogel particles \cite{MD_Neto2016} and MOF microstructures \cite{MD_Tsotsalas2016}. The most common and viable strategy to immobilize water drops is the use of hydrophobic-hydrophilic micropatterns -- the water drops are located on hydrophilic areas bordered by hydrophobic areas. Such chemically patterned substrates are typically produced by top-down lithographic methods including photolithography and microcontact printing \cite{MD_Ueda2012,MD_Butler2001,MD_Zhang2003,MD_Lee2012}. Immobilization of water drops on hydrophobic surfaces has commonly been realized only by means of three-dimensional topographic patterns, such as geometric poly(dimethylsiloxane) (PDMS) obstacles \cite{MD_Huebner2009}, hydrophobic Si pillars \cite{MD_Mandsberg2017}, and superhydrophobic black silicon indentations generated by photolithography and deep reactive ion etching \cite{MD_Yen2015}. 

\begin{figure}[H]
  \centering
  \caption{Skeletal formula of the block copolymer polystyrene-\textit{block}-poly(2-vinylpyridine) (PS-\textit{b}-P2VP).}
  \includegraphics[width=0.4\textwidth]{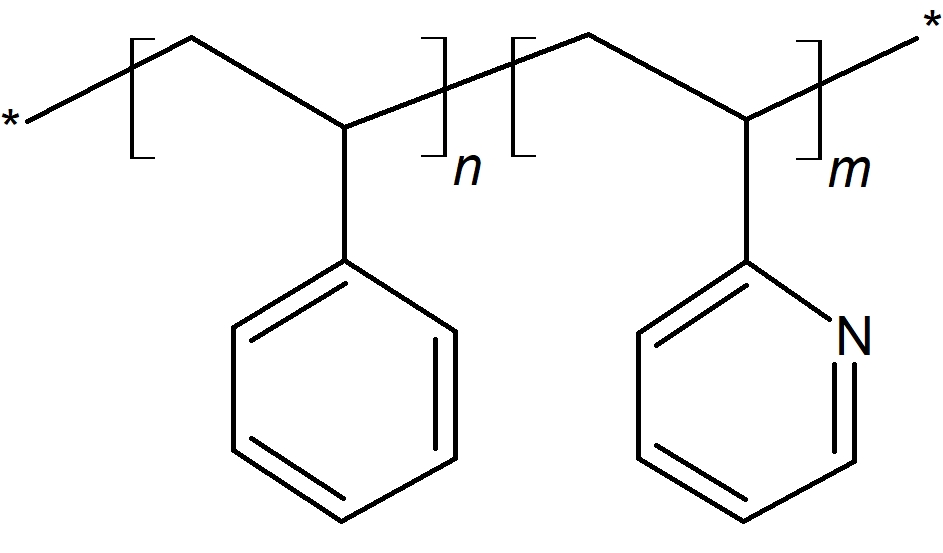}
	\label{PS-b-P2VP}
\end{figure}

\begin{figurehere}
 \begin{center}
  \includegraphics[width=0.7\textwidth]{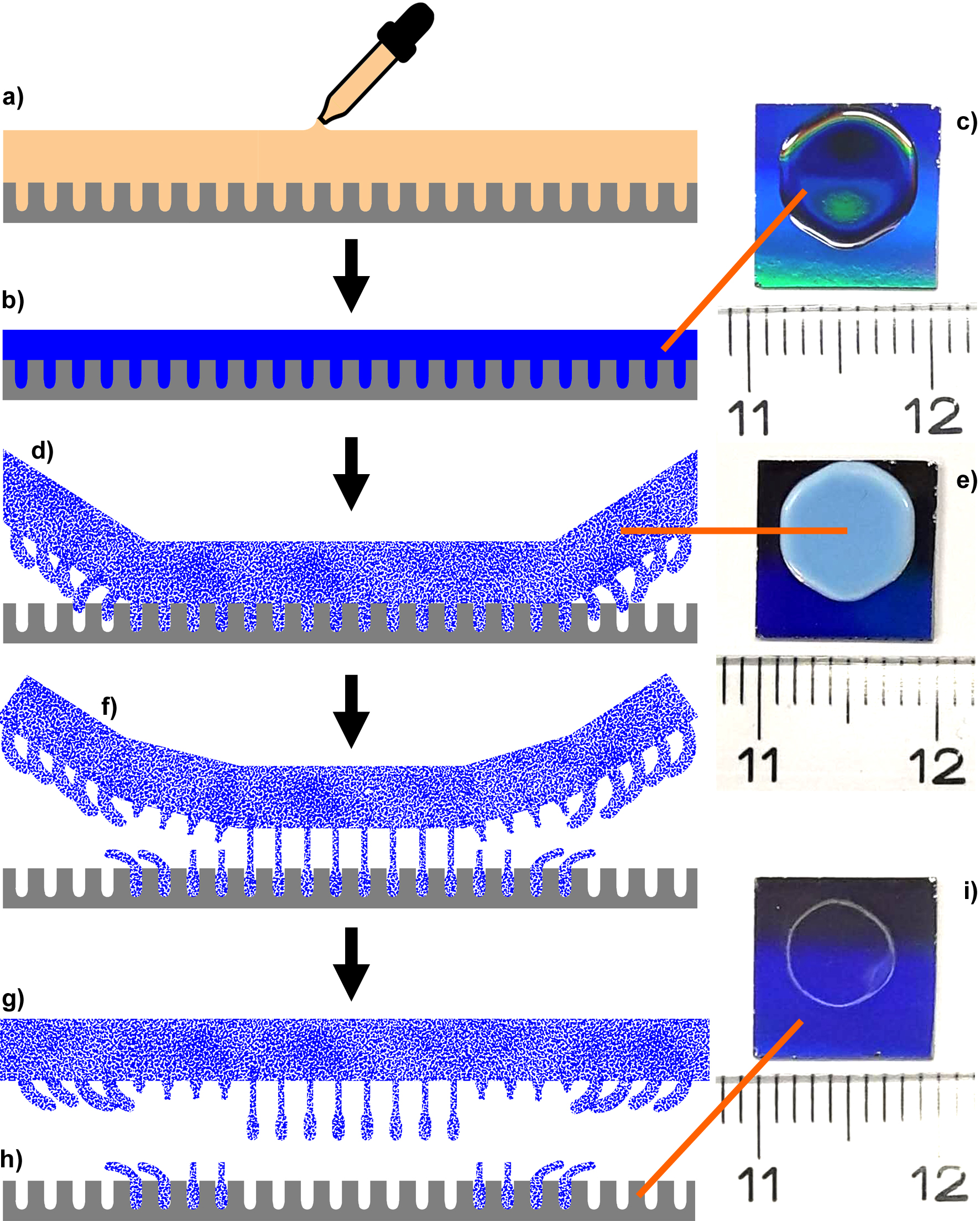}	
  \caption{Nonlithographic generation of PS-\textit{b}-P2VP fiber rings on hydrophobically modified mSi (grey). a) PS-\textit{b}-P2VP/THF solution (ocher) is dropped onto hydrophobically modified mSi so that b), c) a transparent circular film of solid PS-\textit{b}-P2VP (blue) connected to PS-\textit{b}-P2VP fibers located in the mSi macropores forms. d), e) The circular PS-\textit{b}-P2VP film subjected to selective-swelling induced pore formation gets opaque and mesoporous. Volume expansion pushes the PS-\textit{b}-P2VP fibers at the outermost rim out of the mSi macropores. f) The PS-\textit{b}-P2VP fibers in an annular region next to the outermost rim are wedged in the mSi macropores because of the volume expansion associated with selective-swelling induced pore formation. Detachment of the circular PS-\textit{b}-P2VP film results in rupture of these PS-\textit{b}-P2VP fibers, which remain located in the mSi macropores. In the center of the circular PS-\textit{b}-P2VP film the PS-\textit{b}-P2VP fibers are only weakly swollen so that they are completely pulled out of the mSi macropores. g) Detached circular PS-\textit{b}-P2VP film. h), i) After detachment of the circular PS-\textit{b}-P2VP film a ring of ruptured PS-\textit{b}-P2VP fibers located in the mSi macropores remains on the hydrophobically modified mSi.  
\label{scheme}}
 \end{center}
\end{figurehere}

It is highly attractive to immobilize water drops on hydrophobic surfaces that are typically inert, repulsive to adsorbates and characterized by anti-fouling behavior. For example, flexible lab-on-chip configurations, in which water drops are immobilized on hydrophobic rather than on hydrophilic surfaces, could be employed to trap non-adherent cells. However, water drops on hydrophobic surfaces show non-sticking behavior and roll off \cite{MD_Quere2005}. Arresting water drops on hydophobic surfaces has thus remained challenging. Here, we present a nonlithographic approach for the creation of polymeric barriers that efficiently immobilize water drops on highly hydrophobic perfluorinated surfaces. We dropped a solution of asymmetric polystyrene-\textit{block}-polyvinylpyridine (PS-\textit{b}-P2VP) (Figure \ref{PS-b-P2VP}) with P2VP as the minority component onto hydrophobically modified macroporous silicon (mSi) \cite{SI_Lehmann1990,SI_Birner1998} (Figure S1, Supporting Information) and subjected the obtained circular PS-\textit{b}-P2VP film to selective-swelling induced pore formation \cite{SW_Wang2016,SW_Eichler2016a}. Subsequent mechanical detachment of the circular PS-\textit{b}-P2VP film yielded rings of ruptured PS-\textit{b}-P2VP fibers located in the mSi macropores within annular areas having diameters of a few mm and a width of $\sim$0.2 mm (Figure \ref{scheme}). While water drops deposited onto hydrophobically modified mSi without PS-\textit{b}-P2VP fiber rings roll off, the PS-\textit{b}-P2VP fiber rings immobilized water drops having volumes up to 50 $\mu$L even when the hydrophobically modified mSi was turned into a vertical orientation (Figures \ref{20muL} and \ref{60muL}; Supporting Movies 1 and 2). 

\begin{figure}
  \centering
  \caption{Idealized reaction scheme of the grafting of 1H,1H,2H,2H-perfluorodecyltrichlorosilane (PFDTS) onto the native silicon oxide layer covering the mSi macropore walls. As a result, the mSi surface coated by perfluorinated alkyl chains is hydrophobically modified.}
  \includegraphics[width=0.5\textwidth]{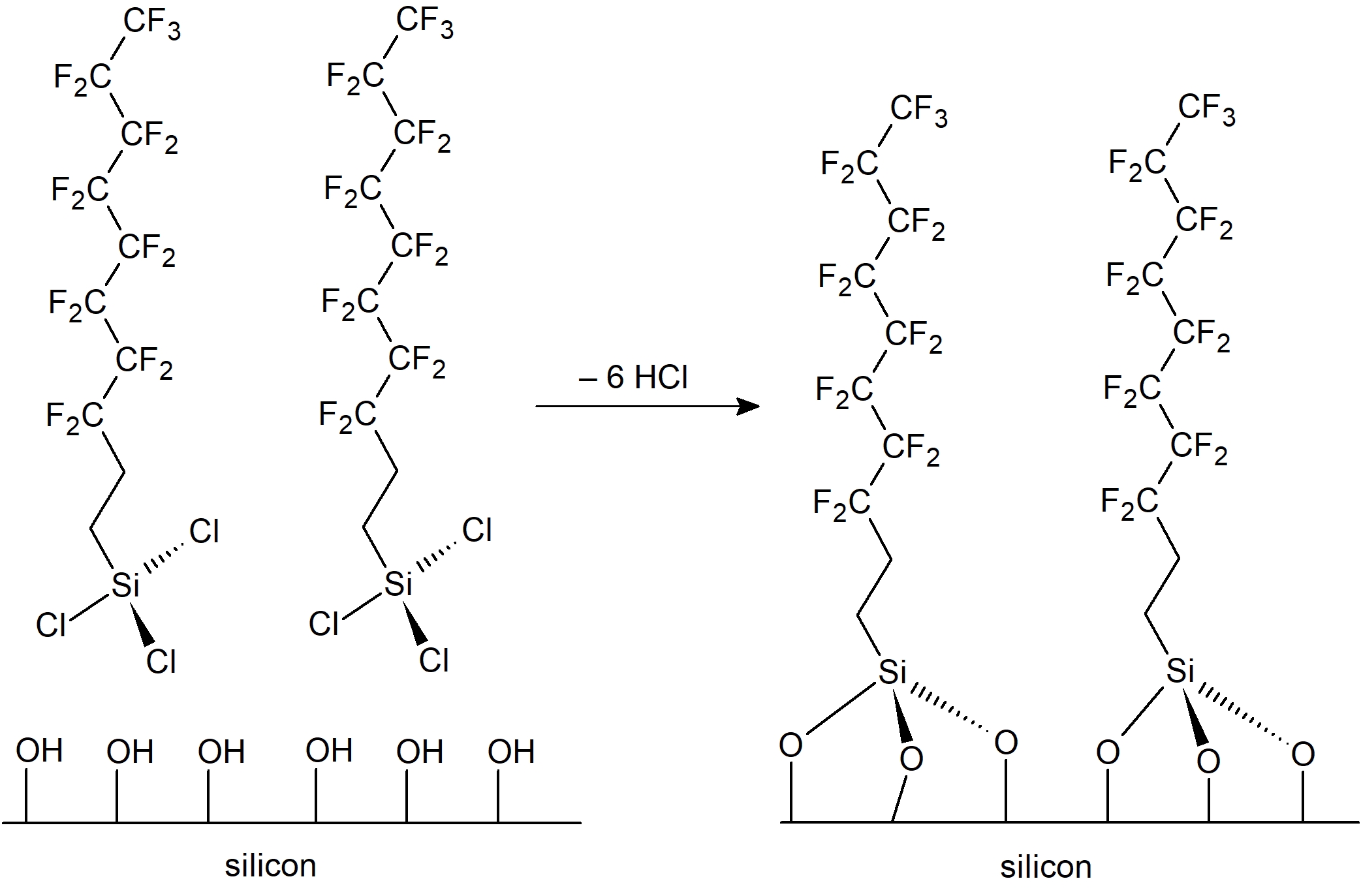}
	\label{PFDTS}
\end{figure}

\section{Results and Discussion}
\subsection{Preparation of PS-\textit{b}-P2VP fiber rings}
The mSi contained macropores with a depth of $\sim$1.8 $\mu$m arranged in hexagonal arrays with a lattice constant of 1.5 $\mu$m (Figure S1a, Supporting Information) and was prepared by a combination of photolithography and photoelectrochemical etching \cite{SI_Lehmann1990,SI_Birner1998}. Below their inverse-pyramidal pore mouths the mSi macropores had necks with a diameter of $\sim$530 nm. Below the necks the mSi macropores widened to $\sim$710 nm (Figure S1b, Supporting Information). We grafted 1H,1H,2H,2H-perfluorodecyltrichlorosilane (PFDTS) onto the native oxide layer covering the mSi following procedures reported elsewhere \cite{SI_Fadeev2000,NP_arrays_Hou2018} (Figure \ref{PFDTS}). In this way, we obtained hydrophobically modified mSi covered by perfluorinated alkyl moieties that exhibited a water contact angle of 127$^\circ \pm$ 4$^\circ$ (average of 6 measurements on different samples; the volume of the deposited water drops was 20 $\mu$L). Dropping a solution containing 0.1 g PS-\textit{b}-P2VP per mL tetrahydrofurane (THF) onto the hydrophobically modified mSi resulted in the formation of transparent, circular PS-\textit{b}-P2VP films. The mSi macropores were filled with PS-\textit{b}-P2VP rods that formed monolithic units with the transparent circular PS-\textit{b}-P2VP films on the hydrophobically modified mSi (Figure \ref{scheme}b and c). In the next step selective-swelling induced pore formation was carried out by treatment with ethanol heated to 60$^\circ$C for 1h. Since ethanol is a solvent selective to P2VP, osmotic pressure drives the ethanol into the P2VP minority domains. The P2VP blocks seek to adopt stretched conformations to maximize their interactions with the ethanol molecules. Hence, the P2VP domains swell and exert pressure on the glassy PS matrix, which undergoes structural reconstruction to accommodate the increasing volume of the swelling P2VP domains. When the ethanol is evaporated, the stretched P2VP blocks undergo entropic relaxation into the coiled state while the reconstructed morphology of the PS matrix is retained. As a result, nanopores form in lieu of the swollen P2VP domains.  

The overall volume of the circular PS-\textit{b}-P2VP films increased in the course of swelling-induced pore formation. However, the PS-\textit{b}-P2VP nanorods located in the mSi macropores impeded lateral expansion of the circular PS-\textit{b}-P2VP films, as obvious from the comparison of Figure \ref{scheme}c and \ref{scheme}e. Therefore, volume expansion nearly exclusively occurred in the direction normal to the mSi surface, as determined from scanning electron microscopy (SEM) images taken at 10 different locations for each sample. The thickness of a circular PS-\textit{b}-P2VP film with a diameter of $\sim$4.4 mm obtained by deposition of 20 $\mu$L PS-\textit{b}-P2VP solution onto hydrophobically modified mSi increased from 111 $\pm$ 1 $\mu$m prior to selective-swelling induced pore formation to 151 $\pm$ 1 $\mu$m after selective-swelling induced pore formation. The thickness of a circular PS-\textit{b}-P2VP film with a diameter of $\sim$6.9 mm obtained by deposition of 60 $\mu$L PS-\textit{b}-P2VP solution onto hydrophobically modified mSi increased from 110 $\mu$m $\pm$ 0 $\mu$m prior to selective-swelling induced pore formation (the standard deviation was rounded to avoid reporting insignificant figures; the rounded standard deviation amounts to 0 $\mu$m) to 147 $\mu$m $\pm$ 1 $\mu$m after selective-swelling induced pore formation. Moreover, selective-swelling induced pore formation altered the appearance of the circular PS-\textit{b}-P2VP films from transparent (Figure \ref{scheme}c) to opaque (Figure \ref{scheme}e). 

The PS-\textit{b}-P2VP fiber rings remaining on the hydrophobically modified mSi have smaller outer contour diameters than the circular PS-\textit{b}-P2VP films they originate from. Dropping 20 $\mu$L PS-\textit{b}-P2VP solution onto hydrophobically modified mSi followed by selective-swelling induced pore formation and detachment of the circular PS-\textit{b}-P2VP films with a diameter of $\sim$4.4 mm yielded PS-\textit{b}-P2VP rings with an outer contour diameter of $\sim$4.1 mm. Dropping 60 $\mu$L PS-\textit{b}-P2VP solution onto hydrophobically modified mSi followed by selective-swelling induced pore formation and detachment of the circular PS-\textit{b}-P2VP films with a diameter of $\sim$6.9 mm yielded PS-\textit{b}-P2VP rings with an outer contour diameter of $\sim$6.5 mm. The outer contour diameter of the PS-\textit{b}-P2VP fiber rings can be controlled by the amount of PS-\textit{b}-P2VP solution deposited onto the hydrophobically modified mSi only within certain limitations. While deposition of 5 $\mu$L or 10 $\mu$L PS-\textit{b}-P2VP solution onto hydrophobically modified mSi yielded circular PS-\textit{b}-P2VP films with diameters of $\sim$2.7 mm and $\sim$3.6 mm, no PS-\textit{b}-P2VP fiber rings remained on the hydrophobically modified mSi after selective-swelling induced pore formation and removal of the circular PS-\textit{b}-P2VP films. Below a certain diameter threshold of the circular PS-\textit{b}-P2VP films annular regions, in which the interplay of swelling and detachment results in the formation of PS-\textit{b}-P2VP fiber rings, do, therefore, no longer form. Instead, the PS-\textit{b}-P2VP fibers are completely pulled out of the macropores of the mSi. Since the hydrophobically modified mSi pieces used in this work had edge lengths of 1 cm, 60 $\mu$L was the largest solution volume that we could deposit. Larger PS-\textit{b}-P2VP fiber rings should, in principle, be accessible. We assume that a soft upper limit to the outer contour diameter of the PS-\textit{b}-P2VP fiber rings might be related to the deposition of the PS-\textit{b}-P2VP solution onto the hydrophobically modified mSi. As the drop volume increases, effects such as contact line instabilities, contact line pinning, or Marangoni convection may become more relevant so that PS-\textit{b}-P2VP films formed on the hydrophobically modified mSi exhibit no longer an approximately circular shapes.

\begin{SCfigure}
  \centering
  \caption{Scanning electron microscopy images of a PS-\textit{b}-P2VP fiber ring obtained by dropping 20 $\mu$L PS-\textit{b}-P2VP solution onto hydrophobically modified mSi. a) Large-field view; the approximate positions at which panels b)-e) were taken are indicated. b), c) Outermost rim at the top (b) and on the bottom (c) of the PS-\textit{b}-P2VP fiber ring shown in panel a). d), e) Center of the PS-\textit{b}-P2VP fiber ring shown in panel a) at the top (d) and on the bottom (e).}
  \includegraphics[width=0.4\textwidth]{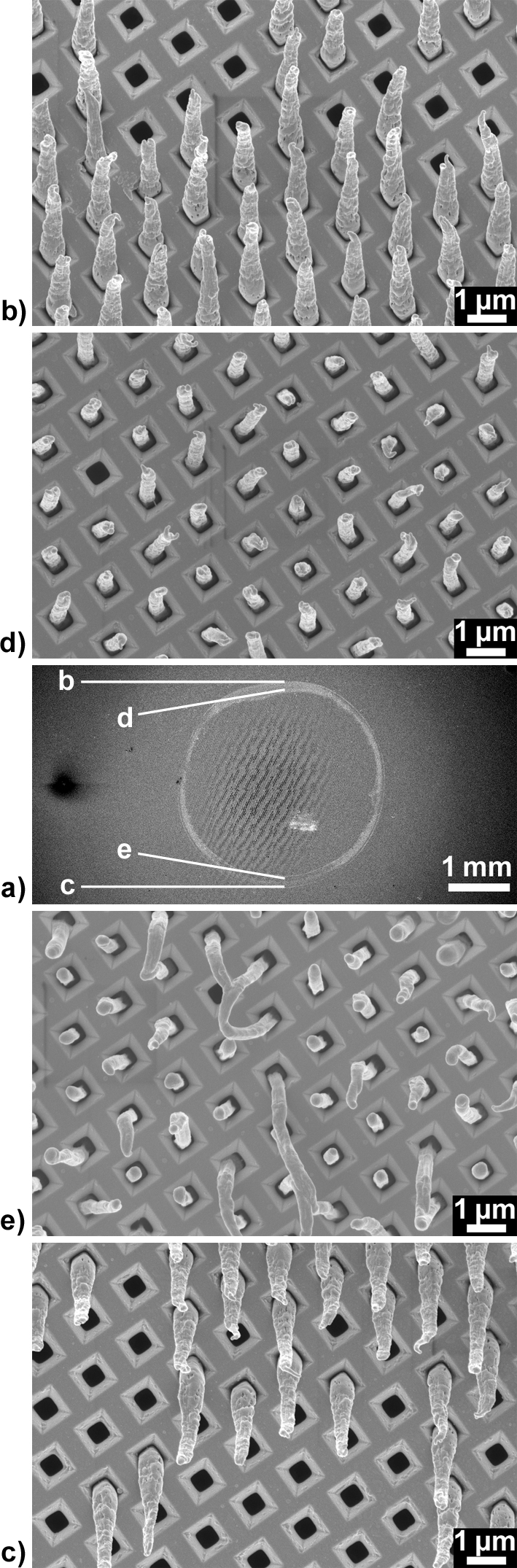}
	\label{SEM}
\end{SCfigure}

\subsection{PS-\textit{b}-P2VP fiber rings: mechanism of formation}
Selective-swelling induced pore formation is crucial for the formation of the PS-\textit{b}-P2VP fiber rings. Figure \ref{SEM}a shows a SEM large-field view of a PS-\textit{b}-P2VP fiber ring obtained by dropping 20 $\mu$L PS-\textit{b}-P2VP solution onto hydrophobically modified mSi (corresponding to Figure \ref{scheme}h,i). Panels b) and c) of Figure \ref{SEM} show details of the outermost rim of the PS-\textit{b}-P2VP fiber ring seen in Figure \ref{SEM}a captured at the top (Figure \ref{SEM}b) and at the opposite position on the bottom (Figure \ref{SEM}c), as marked in Figure \ref{SEM}a. Larger-field SEM views were taken on the left (Figure S2b, Supporting Information), on the right (Figure S2c, Supporting Information), at the top (Figure S2d, Supporting Information) and on the bottom (Figure S2e, Supporting Information) of the outermost rim of the PS-\textit{b}-P2VP fiber ring, as indicated in Figure S2a of the Supporting Information. Within the outermost rim with a width of 10 to 15 $\mu$m the PS-\textit{b}-P2VP fibers protruding from the mSi macropores are bent radially outwards. However, away from the outermost rim the PS-\textit{b}-P2VP fibers protruding from the mSi macropores do no longer show anisotropic orientation (Figure \ref{SEM}d and e; Figure S2f of the Supporting Information). 

Further light on the formation of the PS-\textit{b}-P2VP fiber rings is shed by the investigation of the circular PS-\textit{b}-P2VP films after their detachment from hydrophobically modified mSi (cf. Figure \ref{scheme}g). Figure S3 of the Supporting Information shows a detached circular PS-\textit{b}-P2VP film that is the counterpart to the PS-\textit{b}-P2VP fiber ring displayed in Figure \ref{SEM}a. Figures S4 and S5 of the Supporting Information comprise series of SEM images taken at the positions indicated in Figure S3 of the Supporting Information along a radial line from the top to the bottom of the shown detached circular PS-\textit{b}-P2VP film. The outermost rim of the circular PS-\textit{b}-P2VP film (position U1, Figures S4a and S5a of the Supporting Information; position D1, Figures S4b and S5b of the Supporting Information; position U2, Figures S4c and S5c of the Supporting Information and position D2, Figures S4d and S5d of the Supporting Information) is directly exposed to the swelling agent ethanol. Therefore, selective-swelling induced pore formation starts here and volume expansion pushes the PS-\textit{b}-P2VP fibers out of the mSi macropores. Hence, the outermost rims of the circular PS-\textit{b}-P2VP films detach from hydrophobically modified mSi in such a way that intact PS-\textit{b}-P2VP fibers remain attached to the circular PS-\textit{b}-P2VP films.

As selective-swelling induced pore formation proceeds into the circular PS-\textit{b}-P2VP films, at positions U3 (Figures S4e and S5e, Supporting Information) and D3 (Figures S4f and S5f, Supporting Information) local shear forces pointing radially outwards caused by volume expansion deform the PS-\textit{b}-P2VP fibers. Even though the PS-\textit{b}-P2VP fibers are slightly stretched, they are still completely pushed out of the mSi macropores and remain attached to the circular PS-\textit{b}-P2VP film. Notably, the PS-\textit{b}-P2VP fibers are bent radially inwards.

At positions U4 (Figures S4g and S5g, Supporting Information) and D4 (Figures S4h and S5h, Supporting Information) on the circular PS-\textit{b}-P2VP film a transition from the annular area in which the PS-\textit{b}-P2VP fibers were bent inwards to an annular area closer to the center in which the PS-\textit{b}-P2VP fibers were ruptured occurs. Positions U5 (Figures S4i and S5i, Supporting Information) and D5 (Figures S4j and S5j, Supporting Information) on the circular PS-\textit{b}-P2VP film are located right inside the annular area in which the PS-\textit{b}-P2VP fibers ruptured, which is obviously the counterpart to the PS-\textit{b}-P2VP fiber ring remaining on the hydrophobically modified mSi (Figures \ref{scheme}h,i and \ref{SEM}a). Only the outermost PS-\textit{b}-P2VP fibers of the PS-\textit{b}-P2VP fiber ring protruding from the mSi macropores are bent radially outwards and can be considered as counterparts of the PS-\textit{b}-P2VP fibers bent inwards on the detached circular PS-\textit{b}-P2VP film. The largest portion of the PS-\textit{b}-P2VP fiber ring consists of unoriented PS-\textit{b}-P2VP fibers (Figure \ref{SEM} and Figure S2, Supporting Information). Therefore, radial shear can be ruled out as main origin of the rupture of the PS-\textit{b}-P2VP fibers. Instead, we assume that volume expansion related to selective-swelling induced pore formation becomes effective before the PS-\textit{b}-P2VP fibers are pushed out of the mSi macropores. When the circular PS-\textit{b}-P2VP film is detached from the hydrophobically modified mSi, the PS-\textit{b}-P2VP fibers within the annular areas of the PS-\textit{b}-P2VP fiber rings are wedged in the mSi macropores and rupture. Likely, the neck close to the mouths of the mSi macropores supports the rupturing of the PS-\textit{b}-P2VP fibers. 

Further towards the center of the detached circular PS-\textit{b}-P2VP film, at positions U6 (Figures S4k and S5k, Supporting Information) and D6 (Figures S4l and S5l, Supporting Information) the PS-\textit{b}-P2VP fibers are completely extracted from the mSi macropores and remain, albeit somewhat stretched, attached to the detached circular PS-\textit{b}-P2VP film. As long as the circular PS-\textit{b}-P2VP film is still located on the hydrophobically modified mSi the PS-\textit{b}-P2VP fibers in the center can hardly be reached by the ethanol molecules within the exposure time. Thus, selective-swelling-induced pore formation is impeded, no significant volume expansion occurs, and the PS-\textit{b}-P2VP fibers can be pulled out of the mSi macropores.           

\begin{figure}[htbp]
	     \centering
		\includegraphics[width=1\textwidth]{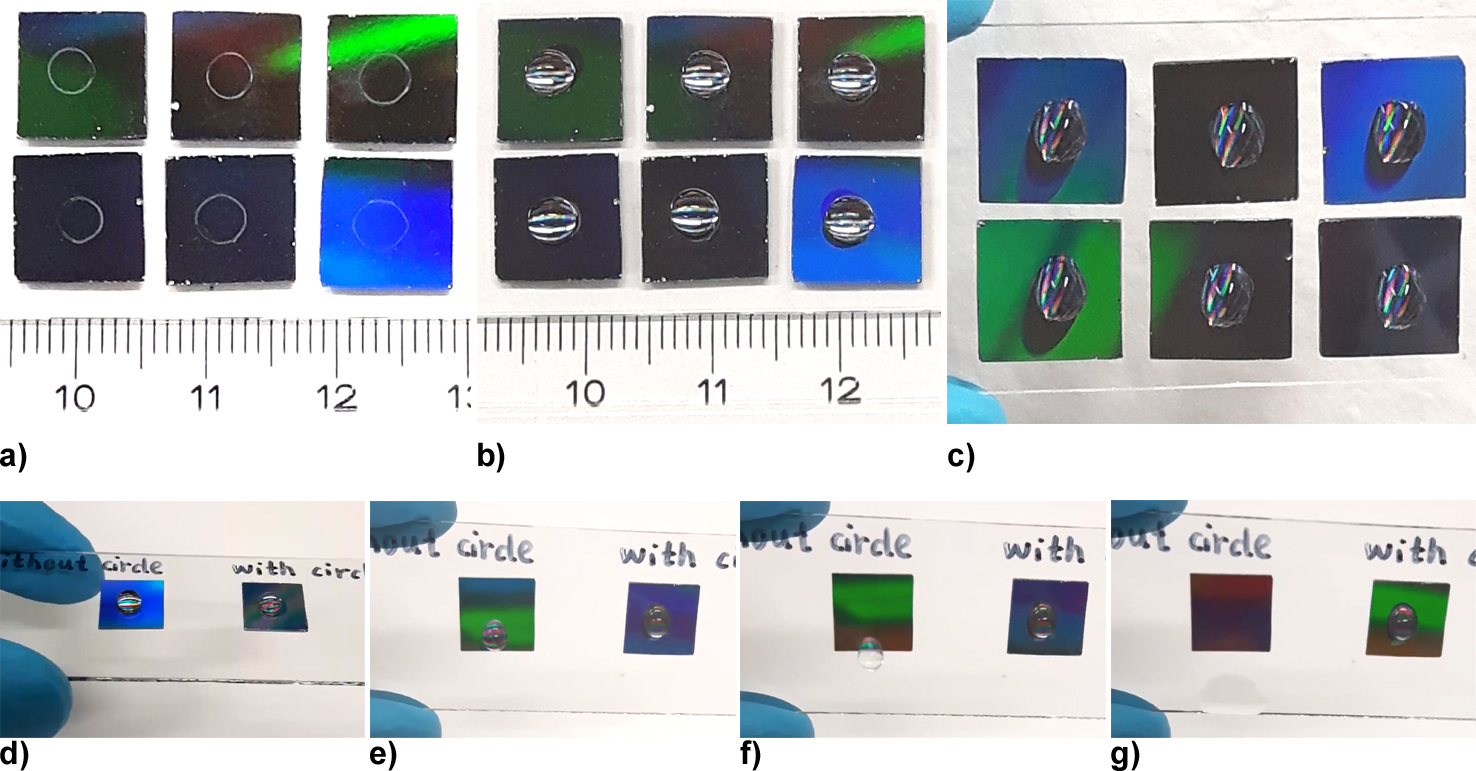}	
	\caption{Immobilization of water drops on hydrophobically modified mSi by PS-\textit{b}-P2VP fiber rings with an outer contour diameter of $\sim$4.1 mm. a) As-prepared PS-\textit{b}-P2VP fiber rings. b) Water drops with a volume of 20 $\mu$L deposited into the PS-\textit{b}-P2VP fiber rings while the mSi substrate is kept in horizontal orientation. c) Water drops with a volume of 20 $\mu$L deposited into the PS-\textit{b}-P2VP fiber rings while the mSi substrate is kept in vertical orientation. d)-e) Comparison of the behavior of water drops with a volume of 20 $\mu$L on hydrophobically modified mSi without (left) and with (right) PS-\textit{b}-P2VP fiber ring. When the hydrophobically modified mSi is tilted from horizontal to vertical orientation, the water drop deposited onto the mSi without PS-\textit{b}-P2VP fiber ring rolls off. The water drop deposited onto the hydrophobically modified mSi with PS-\textit{b}-P2VP fiber ring is held by the PS-\textit{b}-P2VP fiber ring even if the hydrophobically modified mSi is brought into vertical orientation.}
\label{20muL}
\end{figure}

\begin{figure}[htbp]
	     \centering
		\includegraphics[width=1\textwidth]{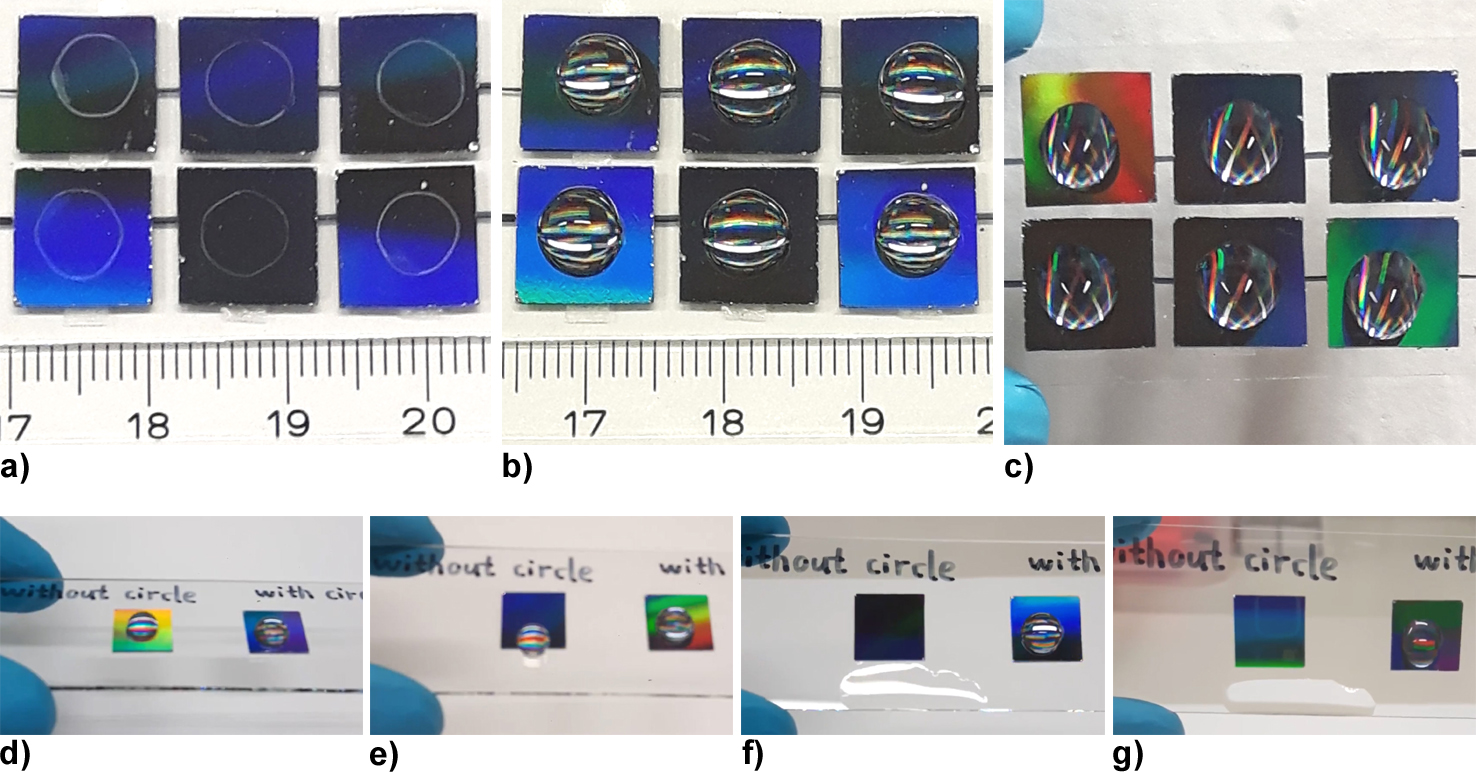}	
	\caption{Immobilization of water drops on hydrophobically modified mSi by PS-\textit{b}-P2VP fiber rings with an outer contour diameter of $\sim$6.5 mm. a) As-prepared PS-\textit{b}-P2VP fiber rings. b) Water drops with a volume of 50 $\mu$L deposited into the PS-\textit{b}-P2VP fiber rings while the mSi sustrate is kept in horizontal orientation. c) Water drops with a volume of 50 $\mu$L deposited into the PS-\textit{b}-P2VP fiber rings while the mSi substrate is kept in vertical orientation. d)-e) Comparison of the behavior of water drops with a volume of 50 $\mu$L on hydrophobically modified mSi without (left) and with (right) PS-\textit{b}-P2VP fiber ring. When the hydrophobically modified mSi is tilted from horizontal to vertical orientation, the water drop deposited onto the mSi without PS-\textit{b}-P2VP fiber ring rolls off. The water drop deposited onto the hydrophobically modified mSi with PS-\textit{b}-P2VP fiber ring is held by the PS-\textit{b}-P2VP fiber ring even if the hydrophobically modified mSi is brought into vertical orientation.}
\label{60muL}
\end{figure}

\subsection{Pinning of water drops on hydrophobically modified mSi by PS-\textit{b}-P2VP fiber rings}
Figure \ref{20muL}a shows six hydrophobically modified mSi pieces with PS-\textit{b}-P2VP fiber rings having outer contour diameters of $\sim$4.1 mm and widths of $\sim$0.2 mm. Water drops with a volume of 20 $\mu$L completely covered the areas encompassed by the PS-\textit{b}-P2VP fiber rings with outer contour diameters of $\sim$4.1 mm and were confined by the latter (Figure \ref{20muL}b). The water drops were even immobilized when the 6 hydrophobically modified mSi pieces were tilted into perpendicular orientation -- despite inertia (when the samples are moved by hand) and gravitation. The photographs shown in Figure \ref{20muL}d-g as well as Supporting Movie 1 highlight the different behavior of water drops with a volume of 20 $\mu$L on hydrophobically modified mSi without (on the left) and with (on the right) a PS-\textit{b}-P2VP fiber ring. As soon as the hydrophobically modified mSi piece without PS-\textit{b}-P2VP fiber ring was tilted, the water drop rolled off; eventually, the water was arrested at the edge of the glass cover slip onto which the hydrophobically modified mSi was glued. However, the water drop deposited into the PS-\textit{b}-P2VP fiber ring on the hydrophobically modified mSi piece seen on the right was arrested even when the hydrophobically modified mSi piece was vertically oriented and despite the trembling movements inevitable when the sample is held manually. Six hydrophobically modified mSi pieces with larger PS-\textit{b}-P2VP fiber rings having outer contour diameters of $\sim$6.5 mm and widths of 0.2 - 0.3 mm  are seen in Figure \ref{60muL}a. Water drops with a volume of 50 $\mu$L completely covered the areas of PS-\textit{b}-P2VP fiber rings with an outer contour diameter of $\sim$6.5 mm (Figure \ref{60muL}b) and remained arrested during tilting into perpendicular orientation (Figure \ref{60muL}c). A drop with a volume of 50 $\mu$L on hydrophobically modified mSi without a PS-\textit{b}-P2VP fiber ring (on the left in panels d-g of Figure \ref{60muL}) rolled off as soon as the hydrophobically modified mSi piece was tilted. The water drop deposited into the PS-\textit{b}-P2VP fiber ring on the hydrophobically modified mSi piece seen on the right in panels d-g of Figure \ref{60muL} was arrested even when the hydrophobically modified mSi piece was vertically oriented (see also Supporting Movie 2).
 
\begin{figure}[htbp]
	     \centering
		\includegraphics[width=0.4\textwidth]{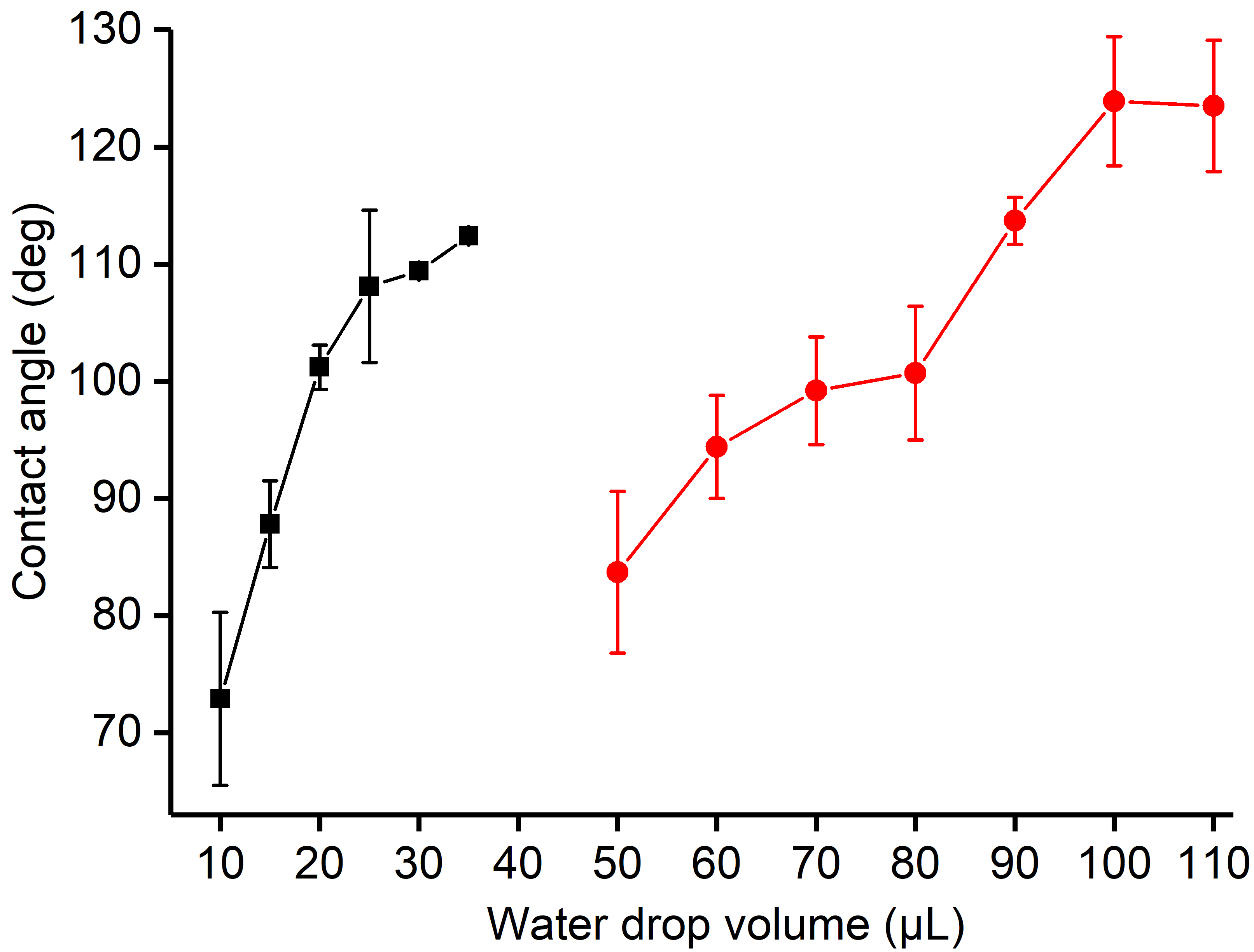}	
	\caption{Apparent contact angles of water drops deposited into PS-\textit{b}-P2VP fiber rings on hydrophobically modified mSi as function of the drop volume. Each data point represents 6 measurements on different samples. Black squares represent data points obtained with PS-\textit{b}-P2VP fiber rings having an outer contour diameter of $\sim$4.1 mm; red circles represent data points obtained with PS-\textit{b}-P2VP fiber rings having an outer contour diameter of $\sim$6.5 mm. The standard deviations are indicated. The lines are guides to the eyes.}
\label{CA}
\end{figure}

Wetting phenomena on topographically patterned substrates are complex \cite{W_Quere2008,W_MacGregor2017} and it has been argued that wetting is controlled by interactions in the vicinity of the contact line, which is the intersection of the liquid phase, the gas phase and the solid substrate \cite{W_Extrand2016}. Water drops with volumes of 5 $\mu$L deposited into PS-\textit{b}-P2VP fiber rings with diameters of $\sim$4.1 mm did neither completely cover the surface area of the hydrophobically modified mSi encircled by the PS-\textit{b}-P2VP fiber rings nor did they contact the latter. The contact angles of the water drops with a volume of 5 $\mu$L deposited into the PS-\textit{b}-P2VP fiber rings amounted to 108$^\circ \pm$ 1$^\circ$. The contact lines of water drops with volumes $\geq$ 10 $\mu$L and $\leq$ 35 $\mu$L coincided with the PS-\textit{b}-P2VP fiber rings. The apparent contact angles increased with increasing drop volume from $\sim$73$^\circ$ $\pm$ 7$^\circ$ to $\sim$112$^\circ$ $\pm$ 0$^\circ$ (Figure \ref{CA}). Water drops with a volume of 40 $\mu$L where no longer confined by PS-\textit{b}-P2VP fiber rings with diameters of 4.1 mm. A similar behavior was observed for PS-\textit{b}-P2VP fiber rings with diameters of $\sim$6.5 mm. The contact lines of water drops with volumes $\geq$ 50 $\mu$L and $\leq$ 110 $\mu$L coincided with the PS-\textit{b}-P2VP fiber rings, while the apparent contact angles increased with increasing drop volume from $\sim$84$^\circ$ $\pm$ 7$^\circ$ to $\sim$124$^\circ$ $\pm$ 6$^\circ$ (Figure \ref{CA}). Water drops with a volume of 20 $\mu$L deposited into PS-\textit{b}-P2VP fiber rings having diameters of $\sim$6.5 mm did not contact the latter and showed contact angles of 123$^\circ$ $\pm$ 1$^\circ$. Hence, water drops within a wide volume range show contact line pinning at the PS-\textit{b}-P2VP fiber rings. The contact line pinning prevents displacements of the water drops out of the PS-\textit{b}-P2VP fiber rings if the hydrophobically modified mSi is moved or tilted but also dewetting within the PS-\textit{b}-P2VP fiber rings. Consequently, water drops may be forced to adopt macroscopic contact angles much smaller than in the absence of the PS-\textit{b}-P2VP fiber rings. The PS-\textit{b}-P2VP fiber rings are both chemical and topographic barriers to contact line displacement. After swelling-induced pore generation the segments of the PS-\textit{b}-P2VP fibers protruding from the mSi pores have an outer surface consisting of P2VP \cite{SW_Wang2008b}. P2VP is slightly hydrophilic and has a water contact angle of 66$^\circ$ $\pm$ 1$^\circ$ \cite{SW_Xue2013}. On the other hand, the PS-\textit{b}-P2VP fibers protruding from the mSi macropores likely pierce into the water drops.



		
			
					
		
    	
				
		


\section{Conclusion}
We have reported the non-lithographic preparation of PS-\textit{b}-P2VP fiber rings with diameters of a few mm and widths of $\sim$0.2 mm on hydrophobically modified macroporous silicon with perfluorinated surface. The rings of ruptured PS-\textit{b}-P2VP fibers protruding from the hydrophobically modified macroporous silicon arrest water drops with volumes up to several 10 $\mu$L by contact line pinning. Thus, displacement of the water drops caused by movements or tilting of the hydrophobically modified macroporous silicon as well as dewetting inside the PS-\textit{b}-P2VP fiber rings are prevented. Water drops deposited onto hydrophobically modified macroporous silicon without PS-\textit{b}-P2VP fiber rings immediately roll off. To prepare the PS-\textit{b}-P2VP fiber rings a PS-\textit{b}-P2VP solution was dropped onto hydrophobically modified mSi. The circular PS-\textit{b}-P2VP films thus obtained were connected with PS-\textit{b}-P2VP fibers located inside the mSi macropores. Subsequent selective-swelling induced pore generation with hot ethanol resulted in expansion of the PS-\textit{b}-P2VP volume. In the outer rims of the circular PS-\textit{b}-P2VP films well accessible to the ethanol molecules swelling-induced pore formation pushed the PS-\textit{b}-P2VP fibers out of the mSi macropores. The centers of the circular PS-\textit{b}-P2VP films with poor accessibility to the ethanol molecules were hardly affected by swelling-induced pore formation; upon detachment of the circular PS-\textit{b}-P2VP films the hardly swollen PS-\textit{b}-P2VP fibers were completely pulled out of the mSi macropores. In the annular regions between the outer rims and the centers of the circular PS-\textit{b}-P2VP films the PS-\textit{b}-P2VP fibers were swollen prior to detachment. Hence, the PS-\textit{b}-2VP fibers were wedged in the mSi macropores and ruptured upon detachment of the circular PS-\textit{b}-P2VP films so that PS-\textit{b}-P2VP fiber rings remained on the hydrophobically modified mSi. The immobilization of water drops on hydrophobic surfaces achieved here enables the exploitation of typical advantageous properties of hydrophobic surfaces such as chemical inertness, anti-fouling behavior and the repulsion of adsorbates in highly flexible lab-on-chip configurations. Water drops immobilized by PS-\textit{b}-P2VP rings on hydrophobically modified mSi may be used as microscale reactores for chemical syntheses or for trapping non-adherent cells. Since mSi can be oxidized to glass prior to hydrophobic modification, integration into advanced optical microscopy set-ups is conceivable. The generation of the PS-\textit{b}-P2VP fiber rings as well as the positioning of water drops into the PS-\textit{b}-P2VP fiber rings can easily be automatized.

\section{Experimental Section}

\textit{Materials.} Macroporous silicon (Supporting Figure S1) was provided by SmartMembranes (Halle, Saale); asymmetric PS-\textit{b}-P2VP (\textit{M}$_n$(PS) = 101000 g/mol; \textit{M}$_n$(P2VP) = 29000 g/mol; \textit{M}$_w$/\textit{M}$_n$(PS-\textit{b}-P2VP) = 1.60, volume fraction of P2VP 21\%; bulk period 51 nm) was obtained from Polymer Source Inc., Canada. 1H,1H,2H,2H-perfluorodecyltrichlorosilane (PFDTS, 97\%, stabilized with copper) was supplied by ABCR GmbH. Tetrahydrofuran (THF), ethanol, 98\% H$_2$SO$_4$, 30\% H$_2$O$_2$ and deionized water were supplied by local manufacturers in the analytical grade.

\textit{Generation of PS-\textit{b}-P2VP fiber rings.} To hydrophobically modify mSi with PFDTS \cite{SI_Fadeev2000,NP_arrays_Hou2018}, mSi was treated with a boiling mixture containing 98\% H$_2$SO$_4$ and 30\% H$_2$O$_2$ at a volume ratio of 7:3 for 30 min, followed by rinsing with deionized water and drying in an argon flow. Then, the mSi was heated for 5 h at 100$^\circ$C in the presence of 0.2 mL PFDTS under room pressure. Defined volumes of a solution containing 0.1 g PS-\textit{b}-P2VP per mL THF were dropped onto the hydrophobically modified mSi using an Eppendorf pipette. The THF was allowed to evaporate for 90 h under ambient conditions. Then, the samples were immersed into ethanol heated to 60$^\circ$ for 1 hour, followed by drying in an argon flow. The circular PS-\textit{b}-P2VP films were detached with tweezers.

\textit{Characterization.} SEM investigations were carried out on a Zeiss Auriga microscope operated at an accelerating voltage of 5 kV. Prior to SEM investigations, the samples were coated with a $\sim$5 nm thick iridium layer. Water contact angles were measured in the sessile drop mode at a humidity of 48.40 \% and a temperature of 23.09$^\circ$C using a drop shape analyzer DSA100 (Kr\"uss, Germany). All contact angle measurements for a specific sample type were repeated on six different samples.\\

\newpage
\textbf{Supporting Information}

Supporting Information is available from the Wiley Online Library or
from the author.\\



\textbf{Conflict of Interest}

The authors declare no competing financial interest.\\


\textbf{Acknowledgement}

The authors thank the European Research Council (ERC-CoG-2014, project 646742 INCANA) for financial support.

\end{document}